\documentclass[useAMS,usenatbib]{mn2e}

\usepackage{graphicx}
\usepackage{subfigure}
\usepackage{color}
\usepackage[authoryear]{natbib}
\usepackage{names}

\begin{document}
\title[]{Dynamical effects of stellar mass loss on a Kuiper-like belt.}

\author[A. Bonsor et al.]
  {A. Bonsor$^1$\thanks{Email: abonsor@ast.cam.ac.uk},
  A. J. Mustill$^1$,
   M. C. Wyatt$^1$\\
  $^1$ Institute of Astronomy, University of Cambridge, Madingley Road,
  Cambridge CB3 0HA, UK}

\maketitle

\begin{abstract}

A quarter of DA white dwarfs are metal polluted, yet elements heavier than helium sink down through the stellar atmosphere on timescales of days. Hence, these white dwarfs must be currently accreting material containing heavy elements. Here, we consider whether the scattering of comets or asteroids from an outer planetary system, following stellar mass loss on the asymptotic giant branch, can reproduce these observations. We use N-body simulations to investigate the effects of stellar mass loss on a simple system consisting of a planetesimal belt whose inner edge is truncated by a planet. Our simulations find that, starting with a planetesimal belt population fitted to the observed main sequence evolution, sufficient mass is scattered into the inner planetary system to explain the inferred heavy element accretion rates. This assumes that some fraction of the mass scattered into the inner planetary system ends up on star-grazing orbits, is tidally disrupted and accreted onto the white dwarf. The simulations also reproduce the observed decrease in accretion rate with cooling age and predict accretion rates in old ($>$1Gyr) white dwarfs, in line with observations. The efficiency we assumed for material scattered into the inner planetary system to end up on star-grazing orbits is based on a Solar-like planetary system, since the simulations show that a single planet is not sufficient. Although the correct level of accretion is reproduced, the simulations predict a higher fraction of accreting white dwarfs than observed. This could indicate that evolved planetary systems are less efficient at scattering bodies onto star-grazing orbits or that dynamical instabilities post-stellar mass loss cause rapid planetesimal belt depletion for a significant fraction of systems.

\end{abstract}


\section{Introduction}


Keck observations of cool single DA white dwarfs find that $\sim$25\% contain elements heavier than helium in their spectra \citep{zuckerman03}. These elements sink rapidly in the white dwarf's atmosphere and their presence means that these white dwarfs must be currently accreting material containing heavy elements. Initially it was thought that these observations were a signature of accretion from the interstellar medium; however, this was ruled out by a lack of correlation between their accreted calcium abundances and spatialâ kinematical distributions relative to interstellar material \citep{Farihi10ism}. The best models to explain these systems \citep{DebesSigurdsson, juraWD03, Gansicke06, Kilic06, vonhippel07, farihi09, farihi10, melis10} suggest that asteroids or comets from the remnants of main sequence planetary systems are scattered onto orbits that approach close to the star, due to altered dynamics following stellar mass loss on the asymptotic giant branch. Bodies that come within the tidal radius of the star are disrupted, potentially forming a dusty disc, before accreting onto the star. Spitzer observations of some of the most highly polluted systems that find excess emission in the near-infra-red, consistent with a close-in dusty disc. Such a disc is observed around 1-3\% of white dwarfs with cooling ages less than 0.5Gyr \citep{farihi09}.

Although the disruption of an asteroid or comet is widely quoted as the explanation for such systems, the feasibility of this process has not been thoroughly investigated. Evidence that the accreted material is asteroidal in nature is high; the composition of the accreted material in systems such as GD40 \citep{klein10} highly resembles asteroids in our solar system. Firstly, in order for this to be the case, planetesimals must survive the star's evolution. Considering only stellar wind drag and sublimation, \cite{jurasmallasteroid} show that asteroids of 1-10km in size survive the giant branch evolution outside of 3-4AU. \cite{bonsor10} model in more detail the evolution of the observed population of debris discs around main sequence A stars, showing that white dwarfs should have planetesimal belts, but that these are hard to detect. 

A mechanism is still required to transport material from an outer planetesimal belt to the star. Although other suggestions, such as stellar encounters (Farihi et al. 2010 in prep), 
have been made, scattering by planets is the most likely mechanism. It is not clear whether planets will survive the star's evolution. Many close-in planets will be swallowed by the expanding stellar envelope whilst the star is on the giant branch (e.g. \cite{villaverlivio}). Multi-planet systems may also become dynamically unstable post-stellar mass loss \citep{DebesSigurdsson}, and planets may be ejected or collide with the star. 

The dynamics of multi-planet systems post-stellar mass loss are complicated. Here, we focus on the effects of stellar mass loss on a single planet and a planetesimal belt. Observations suggest that many debris discs have their inner edges sculpted by planets, similar to the famous example of Fomalhaut \citep{chiang_fom}. Therefore, we consider a planetesimal belt with an interior planet, close enough to the belt such it truncates the inner edge, similar to Neptune and the Kuiper belt in our solar system. The planet dominates the dynamics of bodies at the inner edge of the disc: material inside of the chaotic zone surrounding the planet's orbit \citep{Wisdom1980} will be cleared, due to the overlap of mean motion resonances. As the star loses mass the size of the chaotic zone increases and extra material is scattered from the belt. Here we use N-body simulations to investigate the fate of this scattered material and whether the evolution of this simple system post-stellar mass loss can explain the white dwarf observations. In Sec.~\ref{sec:setup} we describe the set up simulations. In Sec.~\ref{sec:ms} we outline results from our initial simulations that mimic the main sequence evolution of the belt and set up the initial conditions, whilst Sec.~\ref{sec:postms} describes our simulations that include stellar mass loss. In Sec.~\ref{sec:wdobs} we compare our simulations to the white dwarf observations and finally in Sec.~\ref{sec:conc} we conclude and summarise.

\section{Setup }
\label{sec:setup}
In order to investigate the dynamical effects of stellar mass loss on planetesimal belts we consider a simplistic planetary system architecture. Our simulations include a planet and a planetesimal belt orbiting a central star that undergoes mass loss. The main aim is to investigate the fate of the planetesimals in the belt after the star has lost mass. In particular, we consider the feasibility of scattering enough bodies towards the central star in order to produce the hot dusty discs observed around some white dwarfs. 

The simulations are performed using {\it Mercury} \citep{chambers99} with the RADAU integrator. They are set up with a star of mass $M_*$(t), a planet of mass $M_{pl}$ on a circular orbit and N mass-less test particles, in a belt initially outside of the planet's orbit. 
{\it Mercury} was altered such that the central star's mass changes as a function of time. The test particles are distributed in semi-major axis from the planet's semi-major axis $a_{pl}$ to $a_{max}=(\frac{2}{1})^{2/3}a_{pl}$, the 2:1 mean motion resonance, the same outer edge as the Kuiper belt \citep{Trujillo01, Allen01}.

Typically, high mass loss rates on the AGB last for $\sim 10^5$yrs. In all our simulations we consider a 1M$_{\odot}$ star that loses $\frac{2}{3}$ of its mass, at a constant rate, over $10^5$yrs,  however the rate of mass loss and timescale, so long as they are long enough to be adiabatic, will not affect the simulations. Since many particles are removed on short timescales we model the belt expected at the onset of the AGB phase by first running the simulation for the main sequence lifetime, $t_{MS}$ and removing any objects classed as scattered disc (see later), in addition to those ejected or scattered in. Test particles are given randomly selected initial semi-major axis, between $a_{pl}$ and $a_{max}$, eccentricity, between 0 and $e_{max}$, inclination, between 0 and $i_{max}$, mean anomaly, argument of pericentre and longitude of ascending node, between 0 and 2$\pi$. 
Although test particles in these simulations are evenly distributed in semi-major axis, different radial surface density distributions can be considered by appropriate weighting of particles. Each particle is assigned a mass based on its initial semi-major axis and a disc of mass $M_{tot}$, distributed between $a_{pl}$ and  $a_{max}$, with a surface density given by 

\begin{equation}
\Sigma (r) dr \propto r^{-\alpha} dr,
\label{eq:sigma}
\end{equation}
where $\Sigma$ is the mass per unit area, r is the radial distance in the disc and $\alpha$ is a parameter of the models. Since eccentricities are low, a particle's semi-major axis corresponds approximately to its radial position.

During the simulations, on orbits approaching close to the planet, interact with it and are either scattered in towards the central star or out of the system. Some test particles receive a large enough kick that they are put directly onto hyperbolic orbits, become unbound and are ejected. Others undergo a series of scattering interactions increasing their semi-major axis and/or eccentricity. Studies of comets being scattered by a single planet find that when they reach a distance of $a_{galactic}$ from the star, they are more strongly influenced by the galactic tide than the central star \citep{tremaine93}, where
 \begin{equation} 
a_{galactic}=10^4 AU \frac{ (M_{pl}(M_{\oplus}))^{4/3} } {(M_*(M_{\odot}))^{2/3} \, a_{pl}(AU)}.
\end{equation}

  At this point they either become unbound or enter the Oort cloud. In our simulations we assume that a similar process occurs and thus any bodies that go outside of $a_{galactic}$ are classified as `ejected' and removed from the simulation, although occasionally bodies that enter the Oort cloud may return on long period orbits and re-enter the inner planetary system.

The ultimate fate of bodies scattered into the inner system, depends on the planetary system architecture.
As we are considering an arbitrary planetary system this is unknown. Observations of exo-planet systems so far suggest a diversity of architectures. The stability of an arbitrary planetary system post-main sequence is a complicated dynamical question (e.g. \cite{DebesSigurdsson}). In this work our focus is on the dynamics of the planetesimal belt and not the inner planetary system and therefore we merely track particles that are scattered into the inner system, defined as a test particles scattered onto an orbit with $a<a_{in}$, where $a_{in}$ is a parameter of the models. Our working assumption is that a fraction of the bodies that are scattered in will be scattered further times by inner planets and some fraction end up close enough to the white dwarf to be tidally disrupted. In terms of forming the hot white dwarf discs, it is these particles that are of interest.

Test particles that have interacted with the planet, but have not, as yet, been removed we classify as being in the `scattered disc', similar to the Kuiper belt's scattered disc. The scattered disc is defined as test particles with eccentricity higher than $e_{SD}$. Bodies that are in the `scattered disc' will eventually be removed, scattered in or ejected, if the simulation were to run for sufficient time. This means that it is not necessary to run the simulations for the entire main sequence lifetime in order to find the conditions in the belt at the start of the AGB. Although this does mean that the impact of mass loss on bodies in the scattered disc at the end of the main sequence is neglected.

This leaves us with four potential fates for test particles: scattered in (SI), ejected (EJ), in the scattered disc (SD) or left in the belt (B). In this work we track the number of test particles with each of these fates, for a range of simulations in which the different parameters of our model are varied. Parameters we can change are the number of particles, N, planet semi-major axis, $a_{pl}$, planet mass, $M_{pl}$, surface mass density defined by $\alpha$, maximum eccentricity $e_{max}$, maximum test particle inclination, $i_{max}$, the radius inside of which bodies are considered to be scattered in, $a_{in}$, the eccentricity above which bodies are in the scattered disc, $e_{SD}$ and the simulation time before mass loss, $t_{MS}$. We also consider the time evolution on the post-main sequence.

\section{Main Sequence evolution}
\label{sec:ms}
\subsection{Baseline simulation}
\label{sec:baseline}
For our simulation we consider a set-up similar to the solar system's. A Neptune mass planet ($M_{pl}=M_{Nep}$) is placed on a circular orbit at 30AU ($a_{pl}$), with 500 test particles in a belt extending in semi-major axis from 30 to 47.6AU (2:1 resonance). Test particles have initial maximal eccentricities $e_{max}=0.1$ and inclinations $i_{max}=10^\circ$ similar to the cold Kuiper belt. Each test particle was assigned a nominal mass after the simulation was completed based on a disc surface density profile Eq.~\ref{eq:sigma} and $\alpha=1.0$, taken from sub-mm observations of proto-planetary discs (e.g., \cite{andrewswilliams}). Test particles are defined as scattered into the inner system if their semi-major axis is less than $a_{in}$, taken to be $a_{in}= a_{pl} - 7r_{H}$, where $r_{H} = a_{pl} (M_{pl}/3M_*)^{1/3}$ is the Hill's radius, and $7r_{H}$ is half the separation of Neptune and Uranus in our solar system, to the nearest number of Hill's radii. 

In this section we consider this baseline simulation and the effect of changing some of the parameters from this set. Unless explicitly stated all simulations have this set of parameters. The mass that is ejected is defined as $M_{EJ}$, whilst $M_{SI}$ is the mass that is `scattered in' and $M_{SD}$ is the mass that ends up in the scattered disc. The total mass that is scattered, $M_{scatt}= M_{EJ}+M_{SI}+M_{SD}$ and is often quoted as a fraction of the disc mass, $M_{belt}$.

\begin{figure}
\includegraphics[width=80mm] {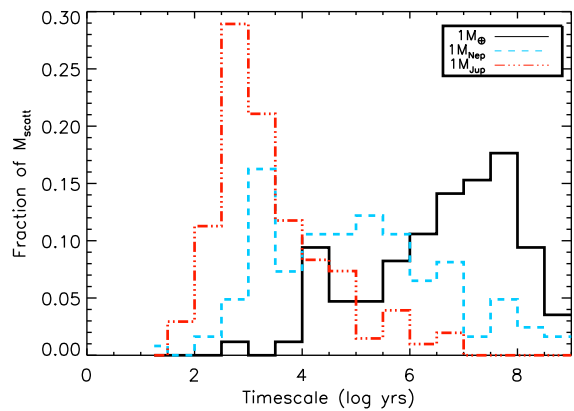}
\caption{The fraction of the total disc mass scattered in (SI) or ejected (EJ) as a function of main sequence simulation time, for the parameters of our initial baseline simulation. }
\label{fig:timescale}
\end{figure}

\subsection{Setting up the initial conditions in the belt}
Before the star undergoes mass loss, the initial conditions in the belt must be set up. This is done by running the simulation and removing all test particles that are ejected, scattered in or end the simulation in the scattered disc. The timescale on which test particles are removed depends on planet mass; more massive planets scatter test particles on shorter timescales. We investigated the timescale on which test particles are removed, shown for several mass planets in Fig.~\ref{fig:timescale}. We found that the number of bodies removed falls off with time and therefore we chose to run our initial simulations for $t_{MS}=10^7$ yrs, as most test particles are removed in this time period. However, it is clear from Fig.~\ref{fig:timescale} that the $1M_{\oplus}$ planet continues to scatter test particles for the entire main sequence (on the order $10^9$ yrs). The initial conditions in the belt will therefore depend on the main sequence lifetime of a particular star. By only using $10^7$yrs to set up the simulation, 45\% of the mass that would end up being removed by a 1M$_{\oplus}$ in $10^9$ yrs is missed, whilst for a 1M$_{Nep}$ this fraction is merely 10\%. This should be taken into account when comparing the results (see later).

\subsection{The effect of varying the definition of `scattered in' or $a_{in}$}

In terms of the formation of the hot white dwarf discs we are interested in the test particles that are `scattered in', assuming that a fraction of these interact with planets in the inner system and are thus scattered onto star-grazing orbits. The fraction that are defined as `scattered in', however, varies significantly with $a_{in}$.

\begin{figure}
\includegraphics[width=80mm] {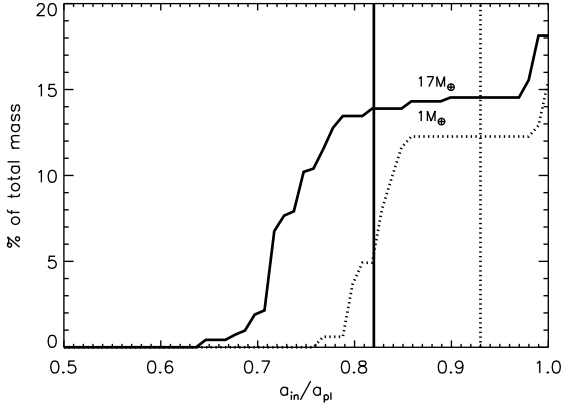}
\caption{The variation in the mass scattered in with changes to the parameter $a_{in}$ for the initial baseline simulation (outlined in Sec.~\ref{sec:baseline}), with a 1$M_{\oplus}$ (dotted), 1$M_{Nep}$ (solid) and e$_{max}=0.1$ and i$_{max}=10^{\circ}$. The dotted and solid vertical lines show $a_{in}=a_{pl} -7r_H$ for 1$M_{\oplus}$ and 1$M_{Nep}$ respectively. }
\label{fig:ain}
\end{figure}

In order to investigate the sensitivity of our results to $a_{in}$ in Fig.~\ref{fig:ain} we changed the definition of $a_{in}$ and calculated the fraction of the total mass that is defined as scattered in (SI). Only a few test particles spend time, at any point during the simulation, just inside of the planet, such that if $a_{in} > 0.98 a_{pl}$ then $M_{SI}\sim 0.04M_{belt}$. These test particles will generally go on to be ejected. $M_{SI}$ is approximately constant for definitions of $a_{in}$ between 0.8 and 0.98. In this region test particles interact strongly with the planet and hence cover the whole range of semi-major axis space. If there are planets in the inner system, interior to $0.8a_{in}$, the dynamics of the test particles will be dominated by interior planets. This behaviour is relatively independent of planet mass or semi-major axis and we expect it scale with the Hill's radius ($r_H$). The assumption of $a_{in}=a_{pl}-7r_H$ in the baseline simulation falls in this region, hence the conclusion of these simulations will be relatively insensitive to $a_{in}$ and if there is another planet at $\sim a_{pl}- 13 r_H$ then interactions with this planet will pull some fraction of planetesimals defined as `scattered in' into the inner region.

These simulations show a very interesting result in terms of the formation of the white dwarf discs. There is a lack of bodies scattered onto star-grazing orbits by a single planet, at 30AU. In fact in these simulations no test particle has a semi-major axis less than 0.5$a_{pl}$. Even including the eccentricity, no test particles has a pericentre of less than 0.3$a_{pl}$. In order for an asteroid to be tidally disrupted and form the observed discs, it must be scattered onto an orbit with pericentre less than the tidal radius (on the order of $R_{\odot}$). It may be that such a small percentage of test particles are scattered further in towards the star, but that our simulations do not find them because we have not included a sufficiently large number of test particles. Given this caveat, these simulations show that a single planet is incapable of producing the observed discs and that an inner planetary system is necessary.

\begin{figure}
\includegraphics[width=80mm] {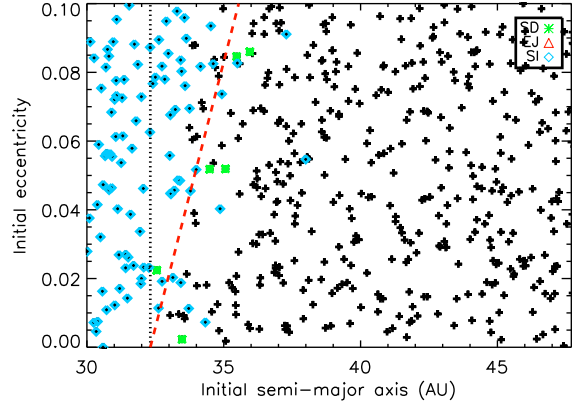}
\caption{The fates of test particles during the initial $t_{MS}$ simulation, as a function of their initial semi-major axis and eccentricity. The parameters take the values for our baseline simulation (outlined in Sec.~\ref{sec:baseline}). Black particles remain in the belt, whilst red particles are ejected, blue scattered into the inner system and green end the $t_{MS}=10^7$yrs in the scattered disc.  The black dot-dashed line shows the size of the chaotic zone (Eq.~\ref{eq:chaos}), whilst the red dashed line shows the region $q<a_{pl}+\delta a_{chaos}$.}
\label{fig:initial}
\end{figure}




\subsection{Comparison to analytic prescription}
An analytic (or semi-analytic) model is useful to understand the physical mechanisms causing instability and allows us to get scaling laws that describe the behaviour over a wide range of parameter space. Analytically it is expected that test particles orbits close to the planet will become chaotic due to the overlap of mean motion resonances. This condition defines the chaotic zone within which this occurs as \citep{Wisdom1980}:
\begin{equation}
\frac{\delta a_{chaos}}{a_{pl}}= C \left (\frac{M_{pl}}{M_*}\right)^{2/7},
\label{eq:chaos}
\end{equation}
where $C=1.3$.

Test particles on chaotic orbits will either be ejected or scattered into the inner planetary system. Thus, the fraction of the disc mass that will be removed, $M_{analytic}/M_{belt}$ can be calculated, for a given surface density profile, assuming that all test particles with initial semimajor axes less than $a_{chaotic} = a_{pl}+ \delta a_{chaos}$ are removed,
\begin{eqnarray}\label{eq:mpre:anal}
M_{analytic}&=& \int^{2\pi}_0 \int^{a_{pl}+\delta a_{chaos}}_{a_{pl}} \Sigma(r) r dr d \theta \nonumber \\
 &=& K C \pi a_{pl}\left( \frac {M_{pl}}{M_*} \right )^{2/7}
\end{eqnarray}

for the index in the surface density profile, $\alpha =1$ and $K=M_{belt}/\pi a_{pl}(2^{2/3}-1)$, for a disc with outer edge, $a_{max}=(\frac{2}{1})^{2/3}a_{pl}$.

This can be compared to our N-body simulations, where we show that most test particles with semi-major axes less than $a_{chaotic}$ are removed, but that some test particles with $a>a_{chaotic}$ are also removed. Fig.~\ref{fig:initial} shows initial semi-major axis and eccentricities of all test particles in the baseline simulation, with those test particles that are scattered by a $1M_{Nep}$ planet highlighted. The higher the initial eccentricity of the test particles the higher the number of test particles outside of the chaotic zone that are removed. This also applies to inclination. No test particles were ejected in the $10^7$ yrs of this simulation as this timescale is too long for a Neptune mass planet to increase a test particles semi-major axis to greater than $a_{galactic}=15,000$AU.

\begin{figure}
\includegraphics[width=9cm] {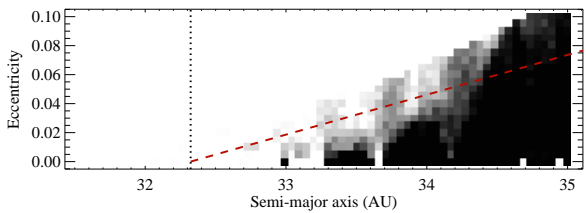}
\caption{The fraction of orbits that are chaotic are a function of initial semi-major axis and eccentricity for orbits around a 1$M_{Nep}$ planet on a circular orbit at 30AU. 100 randomly selected orbits were calculated for each grid point, using an encounter map. The greyscale represents the number of orbits that become chaotic, black representing no chaotic orbits, white all 100 orbits were chaotic. The upper left hand corner of the plot has not been calculated as the encounter map is not valid in this regime. The black dot-dashed line shows the width of the chaotic zone (Eq.~\ref{eq:chaos}), whilst the red dashed line shows the region where test particles have pericentres $q<a_{pl}+\delta a_{chaos}$.}
\label{fig:encounter}
\end{figure}

The formulation for the chaotic zone (Eq.~\ref{eq:chaos}) was developed for bodies on circular orbits. Although \cite{Quillen06} showed that the same formalism applies for eccentric planets, when all test particles have the forced (or the planet's) eccentricity, the behaviour for test particles with high free eccentricities (and inclinations) is different. There are a few sets of simulations that show that the chaotic zone is larger for eccentric or inclined bodies (e.g., \cite{Veras04}), but there is no analytic prescription. Although the formalism for the chaotic zone (Eq.~\ref{eq:chaos}) was developed specifically in terms of semi-major axis, Fig.~\ref{fig:initial} shows that most of the structure in eccentricity can be described by the scattering of test particles with pericentres closer to the planet than the chaotic zone i.e. $q< a_{pl}+\delta a_{chaos}$. This may be because such particles have strong close encounters with the planet, or because of the increased resonance width of eccentric orbits. However, it also seems that the full picture is somewhat more complicated, especially for highly eccentric orbits.

 In order to reproduce the dependence of the chaotic zone width on particle eccentricity, we investigated this behaviour analytically using an encounter map (using the formalism of \cite{Henon86,encounter89}). The encounter map treats the particles as orbiting on unperturbed Keplerian orbits, except at conjunction with the planet where they receive impulsive perturbations. The effect of the perturbations was determined by \cite{encounter89} analytically, and the long-term evolution of an orbit can be determined by repeated application of the analytical formula. In our investigations, the map was iterated for 1,000 synodic periods, corresponding to $\sim1$\,Myr. Orbits were classified as chaotic or regular based on the Fast Fourier Transform of the eccentricity. Regular orbits have smooth Fourier Transforms with a few well-defined peaks at several frequencies. Chaotic orbits have power distributed over a range of frequencies but with large fluctuations in power between closely separated frequencies. To quantify the fluctuations, we defined $P=\log_{10}|\hat{e}|$ where $\hat{e}$ is the FFT of the eccentricity. The fluctuations are then quantified by:
\begin{equation}
N=\left(\Sigma_{i=2}^n(P_i-P_{i-1})^2/(n-1)\right)^{1/2}
\end{equation}
where $n$ is the number of elements in the FFT array. Visual inspection of sample trajectories showed that chaotic orbits typically have $N>0.2$, so this was adopted as the criterion for classifying orbits as chaotic.

 Fig.~\ref{fig:encounter} shows the percentage of orbits that become chaotic for a range of initial eccentricities and semi-major axis, similar to Fig.~\ref{fig:initial}. For each point on the grid 100 orbits, with random initial longitudes of pericentre and mean longitudes, were followed and the black-white scale indicates the percentage of these that are classified as chaotic; black means that 0/100 are chaotic, whilst white means that 100\% are chaotic. This plot reproduces the size of the chaotic zone for zero eccentricity particles, i.e. Eq.~\ref{eq:chaos}, although the factor $C$ is different from the $C=1.3$ given in \cite{Wisdom1980}. However, Fig.~\ref{fig:encounter} shows that a greater fraction of orbits with initially higher eccentricity become chaotic for semi-major axis larger than the chaotic zone. This fits with the behaviour observed in our N-body integrations (see Fig.~\ref{fig:initial}), the higher the initial eccentricities or inclinations, the more particles that are scattered (or end up on chaotic orbits), and can also be approximated by the condition $q<a_{pl}+\delta a_{chaos}$. 


\subsection{Results }

\begin{figure*}
\includegraphics[width=80mm] {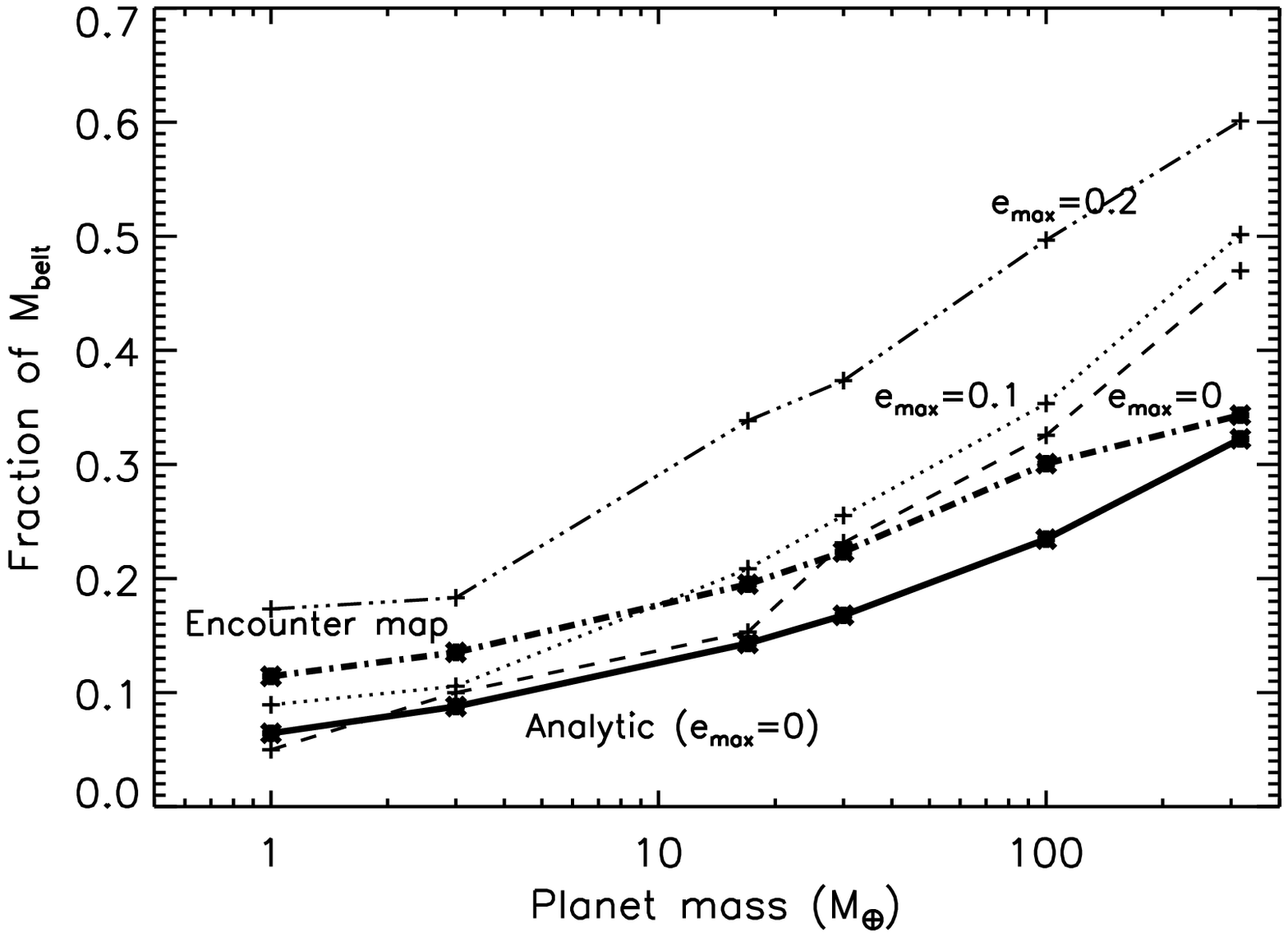}
\includegraphics[width=80mm] {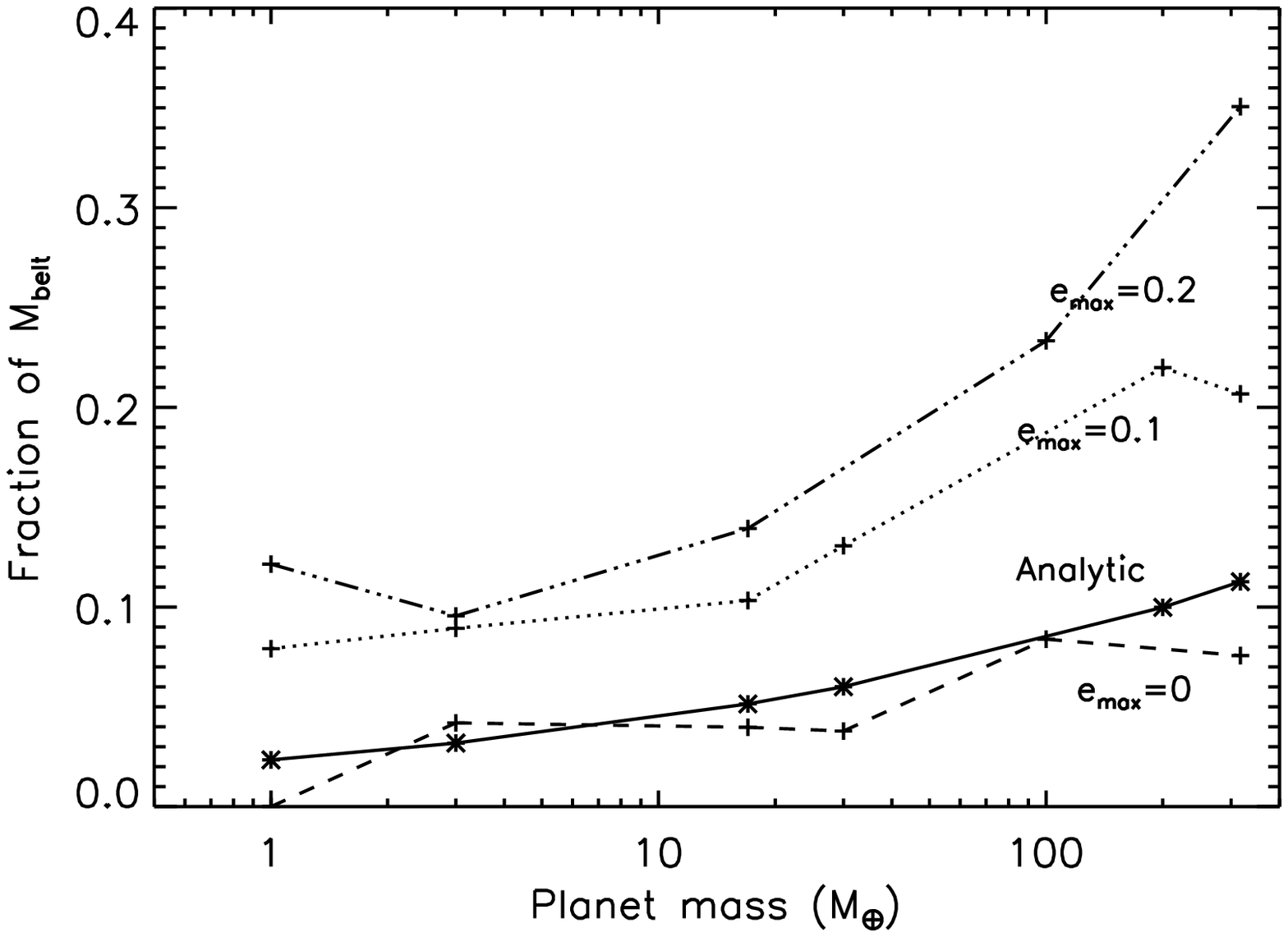}
\caption{The mass removed due to scattering by a planet as a function of planet mass. The numerical simulation with $e_{max}=0.1$ and $i_{max}=10.0^{\circ}$ (dotted) should be compared to the encounter map calculation (dot dashed), whilst the numerical simulation with $e_{max}=0$ and $i_{max}=0^{\circ}$ should be compared to the analytic prescription in which all material within the chaotic zone is scattered (solid). A further numerical simulation with $e_{max}=0.2$ and $i_{max}=20.0^{\circ}$ is shown. 
 The left-hand plot is for the initial $10^7$ yrs simulation without mass loss, whilst the right-hand plot is post-stellar mass loss and 1Gyr of further evolution.   }
\label{fig:prempl}
\end{figure*}

We investigated how the mass scattered by the planet on the main sequence, $M_{scatt}$, varies as a function of planet mass. Our results are shown in Fig.~\ref{fig:prempl} shows the change in the total mass scattered, $M_{scatt}$ (as a fraction of the total disc mass, $M_{belt}$) and a function of planet mass, for the main sequence simulation (i.e. without mass loss). As anticipated, there is an increase in $M_{scatt}/M_{belt}$ with planet mass, for all simulations. 

A numerical simulation with $e_{max}=0.0$ and $i_{max}=0.0$ is compared to an analytic prescription, assuming that all test particles inside of the chaotic zone are scattered (see Eq.~\ref{eq:mpre:anal}). The numerical simulations scatter more mass than predicted analytically. We consider this is due in part to the chaotic zone being larger than predicted by Eq.~\ref{eq:chaos}, for example \cite{chiang_fom} find that $C\sim 2.0$, rather than $C\sim 1.3$, although their simulations are for a mildly eccentric planet. We also find a steeper dependence in $M_{scatt}$ with planet mass than predicted in Eq.~\ref{eq:mpre:anal}. We consider that this is because the simulations were run for the same time ($t_{MS}$), despite the shorter scattering timescales for higher mass planets. If the simulations were run for longer then the lower mass planets would scatter a higher fraction of the disc mass.

The baseline numerical simulation, with $e_{max}=0.1$ and $i_{max}=10.0$ is compared to a prescription calculated using the encounter map. This assumes that particles in the belt are randomly distributed in eccentricity and semi-major axis. 100 particles are placed at every grid point and using the calculation shown in Fig.~\ref{fig:encounter}, the fraction of the total disc mass in bodies not on chaotic orbits is calculated, for the given surface density profile. $M_{scatt}$ calculated using the encounter map prescription is somewhat lower than the numerical simulations, presumably because the chaotic zone increases in size with inclination as well as eccentricity, but inclined particles were not included in the encounter map. We conclude that lower mass planets scatter a smaller fraction of the disc mass due to the longer scattering timescales, as discussed above.

 Analytically it is anticipated that these results are independent of $a_{pl}$, which was verified by simulations, however changing the surface density profile changes $M_{scatt}$ by up to 15\%, for $0.8<\alpha <2.0$. 

These simulations show that there is a definite increase in $M_{scatt}$ with initial eccentricity and inclination. Of the values of initial eccentricity and inclination tested here, the numerical simulation with $e_{max}=0.2$ and $i_{max}=20^\circ$ scatters the most test particles. This implies that an analytic formulation for the chaotic zone for inclined or eccentric particles should be larger than that given in Eq.~\ref{eq:chaos}. Thus, the expected disc structure at the end of the main sequence evolution will be the initial disc, minus material originally inside of the chaotic zone, that was scattered during the initial simulation.

\section{Post-main sequence evolution}
\label{sec:postms}
\subsection{Analytic formulation}

Once the orbital distribution at the end of the main sequence has been determined using the simulations of Sec.~\ref{sec:ms}, we then studied evolution beyond the main sequence, including stellar mass loss. The star loses mass on timescales that are long compared to the orbital timescales and thus this should be an adiabatic process. Indeed this is seen to be the case for all test particles not scattered by the planet and the planet itself. As the star's mass decreases by a factor of 3, their orbital radii increase by the same factor, whilst their eccentricities and inclinations remain constant. This would happen for all particles and the planet itself; however as the stellar mass decreases and the ratio of the planet's mass to the stellar mass increases, the zone of influence of the planet increases. For these simulations, where the stellar mass is decreased by a factor of 3, the size of the chaotic zone increases by a factor of $3^{2/7}$ (see Eq.~\ref{eq:chaos}). Analytically a prediction for the amount of mass scattered can be found by assuming that all test particles inside of the chaotic zone post-mass loss, but outside of its smaller pre-mass loss value, are scattered, given by:
\begin{eqnarray}\label{eq:manal}
M_{analytic}&= & \int^{2\pi}_{0} \int^{[a_{pl}(0)+ 3^{2/7}\delta a_{chaos}(0)]}_{[a_{pl}(0)+ \delta a_{chaos}(0)]} \Sigma(r) r dr d \theta \nonumber \\
&=& K (3^{2/7} -1) \delta a_{chaos}(0), 
\end{eqnarray}
where $\delta a_{chaos}(0)$ is the initial size of the chaotic zone (Eq.~\ref{eq:chaos}), $a_{pl}(0)$ is the planet's initial semi-major axis, $\alpha$ is taken as 1.0 and $K=M_{belt}/\pi a_{pl}(0)(2^{2/3}-1)$ is a constant determined from the initial belt mass.

 The right-hand panel of Fig.~\ref{fig:prempl} compares the amount of mass scattered, following mass loss and 1Gyr of further evolution, found in the numerical simulations, to the analytic increase in the size of the chaotic zone. The numerical simulations show approximately the same dependence with planet mass as the analytic prescription. The simulation with $e_{max}=0.0$ and $i_{max}=0.0$ is closest to the analytic prescription, whilst as anticipated the simulations with higher initial eccentricities and inclinations scatter more test particles.

The main cause of scattering post stellar mass loss in these simulations is the increase in the extent of the chaotic region close to the planet. This extent can be estimated analytically from Eq.~\ref{eq:chaos}, giving Eq.~\ref{eq:manal}, however this underestimates $M_{scatt}$ by a factor of a few if the belt is initially dynamically hot.

\begin{figure*}
\includegraphics[width=80mm] {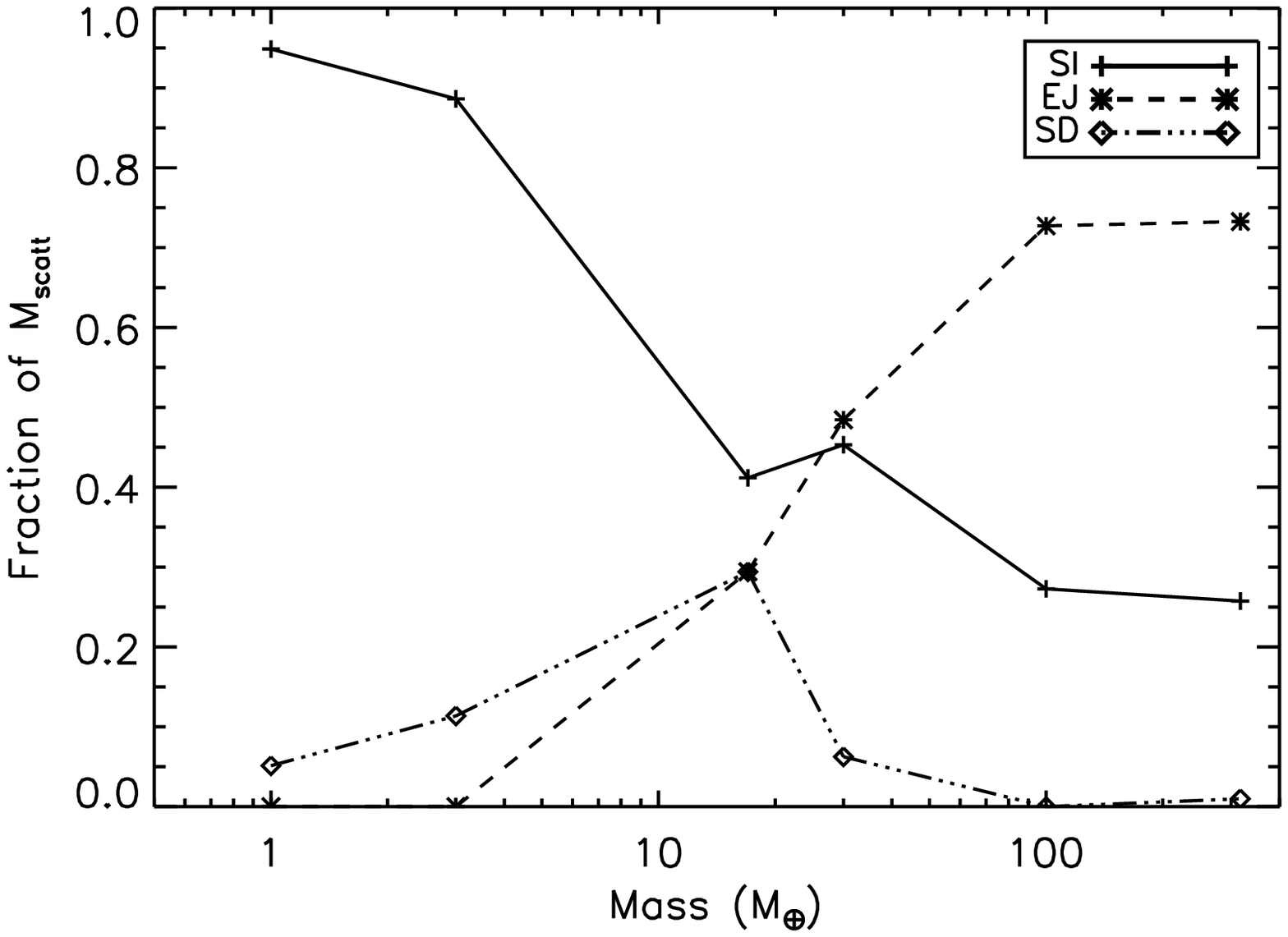}
\includegraphics[width=80mm] {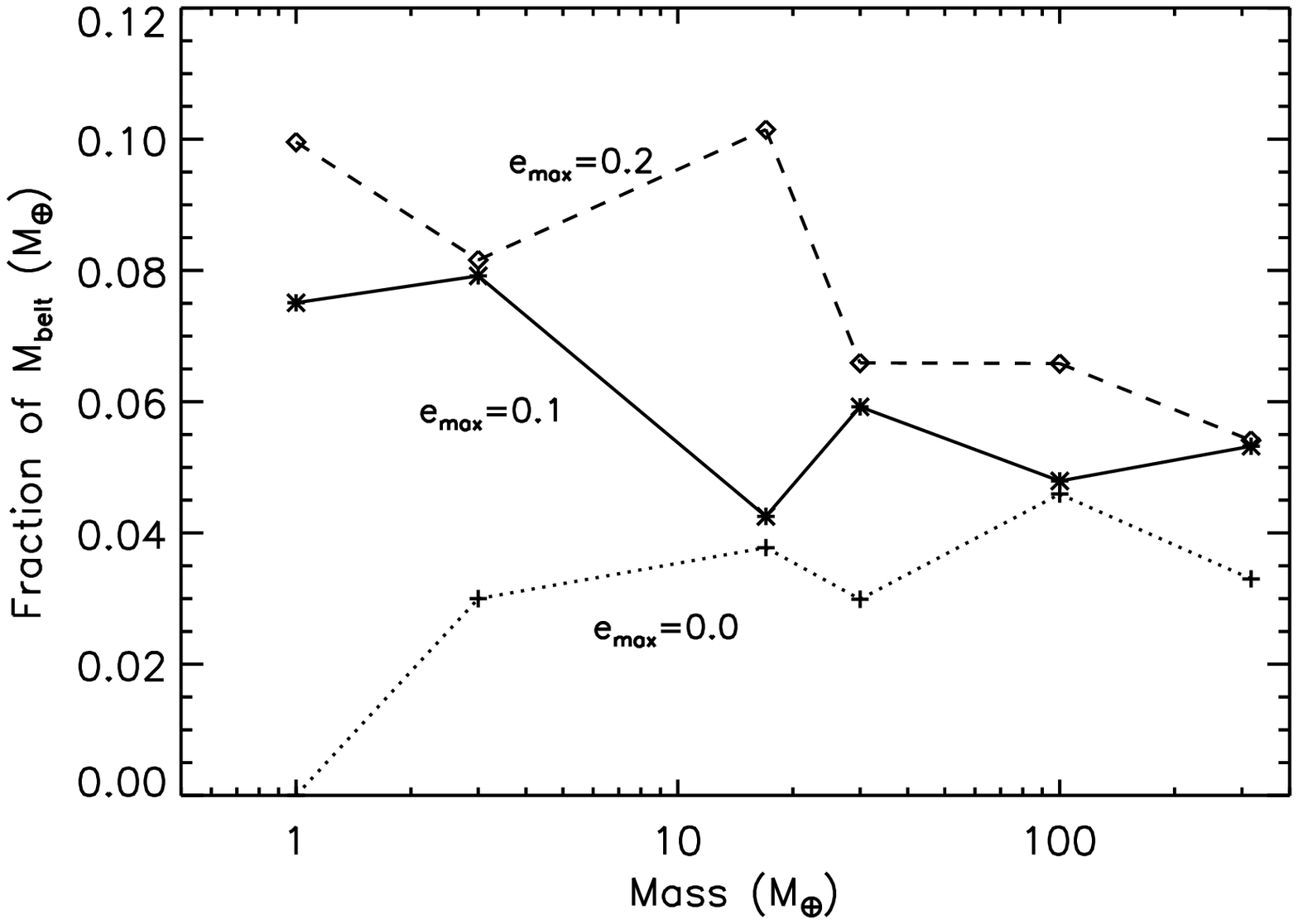}
\caption{The left hand panel shows $M_{SI}$, $M_{EJ}$ and $M_{SD}$ as a fraction of $M_{scatt}$, for various mass planets. The remainder of $M_{scatt}$ ends the simulation in the scattered disc. The right hand panel shows $M_{SI}$ as a function of planet mass for different initial eccentricities and inclinations. }
\label{fig:mpl}
\end{figure*}



\subsection{Scattered in or ejected?}
We investigated the fate of scattered bodies as a function of planet mass. This is shown in the left hand panel of Fig.~\ref{fig:mpl}. The analytic formulation does not give the ultimate fate of bodies, so it is necessary to use N-body simulations. The majority of the mass that is scattered by the planet ends up in the inner planetary system, according to our definition of `scattered in'. The fraction of the mass that is ejected increases with planet mass. This is because higher mass planets give test particles a much larger kick per encounter and thus are more likely to scatter bodies out of the system after fewer encounters. Multiple encounters are required however to raise the test particle's eccentricity high enough that it is ejected. Assuming that the Tisserand parameter is conserved, for a test particle encountering the planet with a semi-major axis equal to the planet's and zero inclination, the condition $e>0.405$ must be satisfied for it to be ejected. No test particles in our simulations start with such a high eccentricity and hence several encounters, each of which increase the eccentricity are required before a particle is ejected.

 Lower mass planets, on the other hand, tend to scatter test particles many many times before they get ejected. This leads to an increased chance of a test particle having $a<a_{in}$ at some point before it is ejected and hence in our definition being scattered in. There is a general trend that higher mass planets scatter test particles on shorter timescales and therefore clear any bodies from the scattered disc before the end of the simulation, whereas lower mass planets end the simulations with a higher mass in the scattered disc. However, the lowest mass planets scatter test particles more weakly and thus fewer test particles end up with $e>e_{SD}=0.24$, our definition of the scattered disc and $M_{SD}$ is low for the lowest mass planets.



Interestingly although $M_{scatt}$ is very dependent on planet mass, the dependence with planet mass on $M_{SI}$ is weak, at least for $e_{max}=0.1$ and $i_{max}=10^{\circ}$; see the right hand panel of Fig.~\ref{fig:mpl}. This is because the decrease in $M_{SI}/M_{scatt}$ with planet mass counteracts the increase in $M_{scatt}$. $M_{SI}$ is highest for discs with initially high eccentricities and inclinations. This is because $M_{scatt}$ is also higher since the size of the chaotic zone increases with eccentricity/inclination (see explanation above). Further simulations also show that $M_{SI}$ is also independent of the planet's semi-major axis.







\begin{figure*}
\includegraphics[width=80mm] {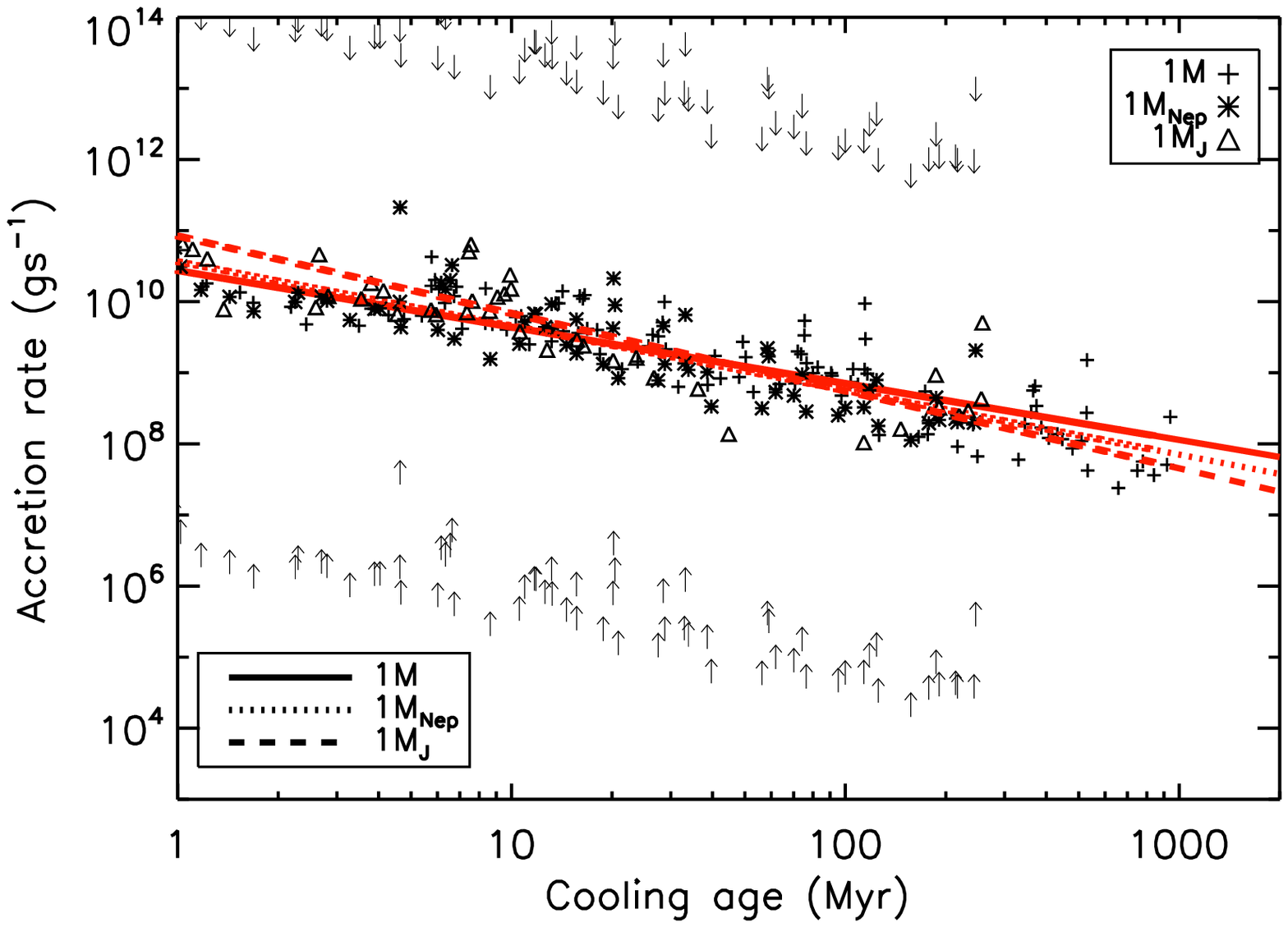}
\includegraphics[width=80mm] {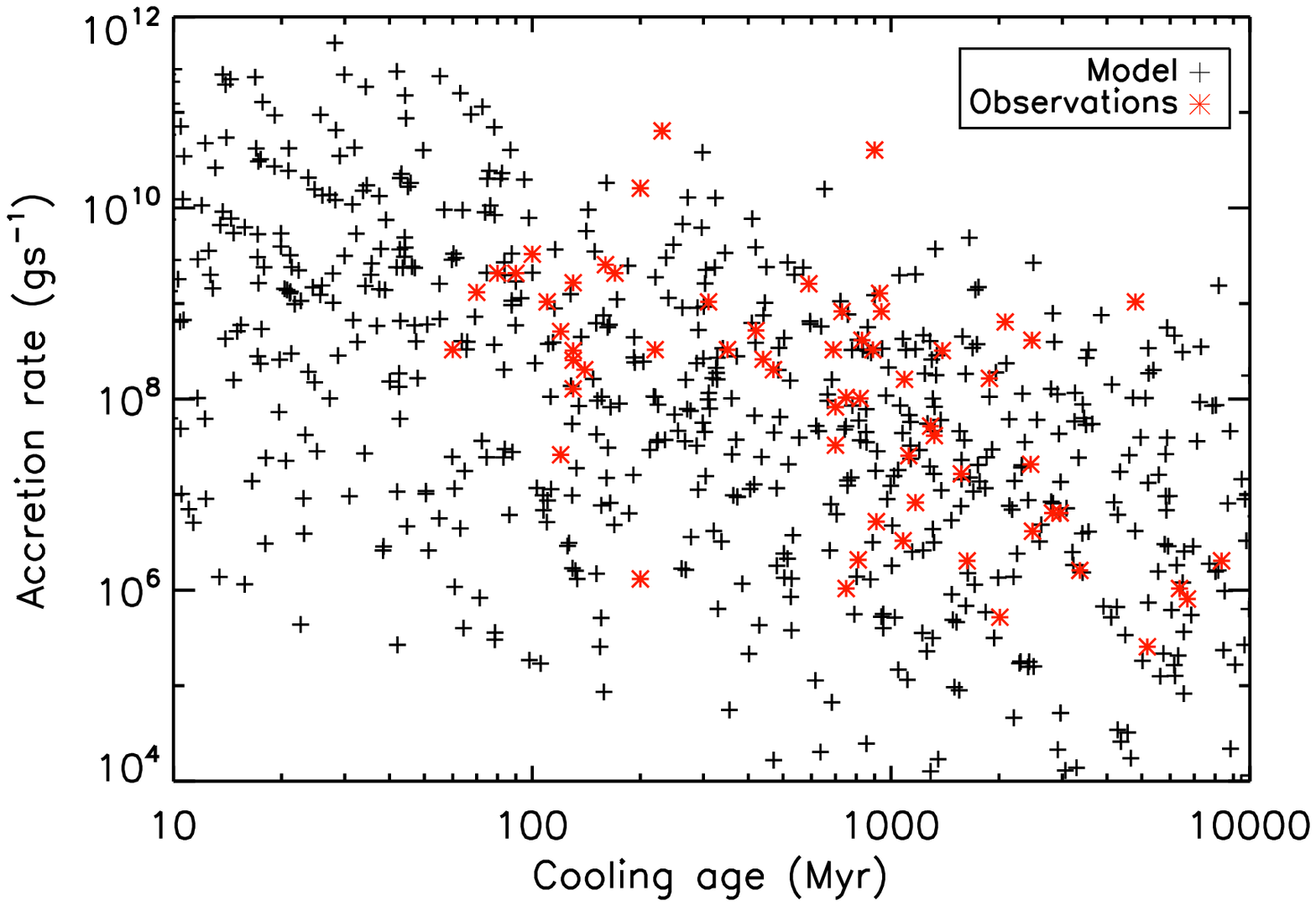}
\caption{Left panel:The accretion rate calculated using Eq.~\ref{eq:m} and Eq.~\ref{eq:scatt} for our simulations, with $M_{pl}=1M_{\oplus}$ (crosses and solid line), 1$M_{Nep}$ (asterisks and dotted line) and 1$M_J$ (triangles and dashed line). A range of belt masses are calculated using the population models of \citep{bonsor10}, but only the median value is plotted here, with upper and lower limits for the 1$M_{Nep}$ case shown by the arrows. Each test particle that is scattered in is plotted with a discrete value of $\dot M$. A straight line is then fitted to these data points.
Right panel: The accretion rates for a population of discs with randomly selected initial belt mass, radius, cooling age and stellar properties, using the smoothed fit to the stochastic accretion process, as determined in the left-hand panel, but for each individual disc mass. These are compared to observed heavy element accretion rates from \citet{farihi09} (red asterisks).} 
 \label{fig:posttime}
\end{figure*}

\begin{figure}
\includegraphics[width=80mm] {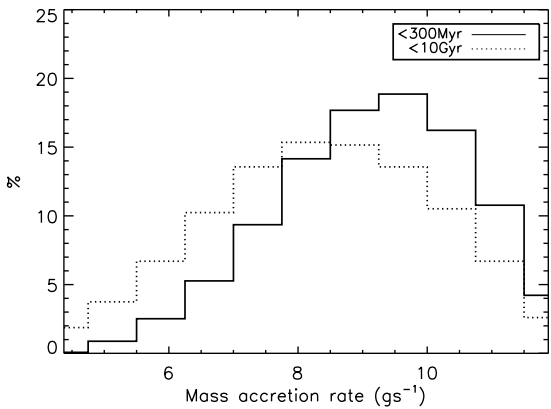}
\caption{A histogram of the predicted accretion rates from our baseline simulation, taken from the population of discs in Fig.~\ref{fig:posttime}. Discs around white dwarfs with cooling ages between 100Myr and 300Myr, and less than 10Gyr are shown. }
\label{fig:mhist}
\end{figure}

\section{The relationship between these simulations and observations of metal rich white dwarfs.}
\label{sec:wdobs}

In relation to the observations of metal polluted white dwarfs and/or white dwarfs with dusty discs, we are interested in bodies that are scattered into the inner planetary system. We assume that some fraction, $f_{TD}$, of the material scattered into the inner planetary system is further scattered by interior planets and ends up on an orbit that approaches close enough to the star for the body to be tidally disrupted. It is these bodies that potentially form the observed discs and accrete onto the star. This fraction is highly dependent on the inner planetary system architecture and will vary between individual planetary systems. Observations of exo-planet systems so far suggest a diversity of architectures. The stability of an arbitrary planetary system post-main sequence is a complicated dynamical question (e.g. \cite{DebesSigurdsson}) and it is therefore beyond the scope of the current work to investigate in detail scattering by the inner planetary system. 


$f_{TD}$ can be approximated for our solar system from previous simulations. \cite{LevisonDuncan97} find that $\sim$1\% of visible comets end up on sun-grazing orbits. The particles in their simulations that leave the Kuiper belt to become visible comets correspond approximately to our definition of `scattered in'. \cite{LevisonDuncan94}, on the other hand, investigate the fate of known visible comets in our solar system and find that $\sim$6\% end up on sungrazing orbits. The discrepancy between the two is likely to be due to the inclusion of the terrestial planets in \cite{LevisonDuncan94}. Therefore, we adopt a fraction $f_{TD}=0.06$ of $M_{SI}$ that ends up on sun-grazing orbits and is therefore tidally disrupted. Of course, $f_{TD}$ should be calculated for each individual planetary system and could vary between 0 and 1. 

Not all of the material that is scattered in close enough to the star to be disrupted will end up in a disc, or be accreted onto the star. The formation of a such disc has not been modeled in detail at present, but here we assume that the disruption is relatively inefficient and only a small fraction, $f_{acc}\sim 0.1$, of the mass that reaches $\sim R_{\odot}$, ends up accreting onto the star. Hence, the mass that will be accreted onto the star is given by:
\begin{equation}
M_{acc} \sim f_{acc} \times f_{TD} \times f_{SI} \times M_{belt} ,
\label{eq:m}
\end{equation}
where $f_{SI}=M_{SI}/M_{belt}$ is the fraction of the initial belt mass defined as `scattered in'. $M_{SI}$ can either be the total mass that is scattered in, and then $M_{acc}$ is the total mass that is accreted over the white dwarf lifetime, or alternatively the mass scattered in within a time interval $dt$, in which case $M_{acc}$ is the mass accreted in the time interval $dt$.

Spitzer near-infrared observations of white dwarfs are used to determine dust masses for the observed discs. Since discs are opaque, this is a minimum disc mass and it is unclear how it relates to the total disc mass or the mass that must be disrupted into order to produce such an observation. We, therefore, chose to compare the results of our simulations to the heavy element accretion rates calculated from observed abundances of metals in the white dwarf atmosphere. These are calculated from observed Ca abundances, an assumption that the accretion is in steady state and that the abundance of Ca in the accreting material is approximately solar. 

Assuming that mass must be supplied to the disc at least at the rate at which it accretes onto star, the results of these simulations can be interpreted in terms of the observations. 
The rate at which mass is scattered inwards onto star-grazing orbits, or the predicted accretion rate, is given by:
\begin{equation}
\dot M_{acc} \sim \frac{M_{acc}}{ \Delta t},
\label{eq:scatt}
\end{equation}
where $\Delta t$ is the time interval over which a mass $M_{acc}$ is scattered. This assumes that the accretion is a continuous process and that the accretion rate is determined by the scattering rate rather than viscous timescales in the close-in disc. These accretion rates could, however, be considered a minimum for the rate at which material must be supplied to the disc in order to reproduce the observed heavy element accretion rates onto the star. If the pollution is produced by the disruption of a large individual body, as suggested by, amongst others, \cite{juraxray, debesasteroidbelt} then a lower scattering rate than predicted by these simulations is required.

Using this formulation and these assumptions, we calculated the accretion rate from each individual scattering event. The timescale for scattering, $dt$ in Eq.~\ref{eq:scatt} is calculated as the mean of the time between the current scattering event and those immediately preceding and following it. Properties of the disc were selected randomly from the main sequence population of debris discs around A stars, and the collisionally evolved mass at the end of the main sequence determined, according to the models of \cite{wyatt07} and \cite{bonsor10}. The mass left in the disc after our initial simulations was equated with the collisionally evolved mass at the end of the main sequence. Collisional evolution of the disc mass in the white dwarf phase is negligible \citep{bonsor10}. The left-hand panel of Fig.~\ref{fig:posttime} shows these accretion rates as a function of time, for a belt truncated by a 1$M_{\oplus}$ (crosses), 1$M_{Nep}$ (asterisks) and 1$M_J$ (triangles) planet. Only the belt with the median mass at the end of the main sequence is displayed, whilst the arrows show the highest and lowest mass belt in the population, for the 1$M_{Nep}$ case.

 As anticipated, early in the white dwarf phase many particles are scattered, whilst at later times, the number of particles scattered as a function of time, and thus the accretion rate, decreases. This happens slightly more slowly for the lower mass planets, since scattering times decrease with increasing planet mass. The difference between the different planet masses, however, is small, compared to the range of accretion rates for different initial belt properties, or the other assumptions that went into this plot. In order to convert this stochastic process into a smooth decrease with time, a straight line was fitted to the data for each belt mass. These are shown for the belt with median mass by the solid (1$M_{\oplus}$), dotted (1$M_{Nep}$) and dashed (1$M_J$) lines. For a 1$M_{Nep}$ planet the slope of this line is $-1.1 \pm 0.04$. Observations of metal polluted white dwarfs also show a decrease in accretion rate with cooling age of the white dwarf (e.g. \citep{farihi09}). Using a sample of 62 white dwarfs from \cite{farihi09, farihi10} we found that the decrease in $\log\,(t_{cool})$ of $\log\,({\dot M})$ can be fitted with a straight line of slope of $-1.3 \pm 0.23$. This compares well with our simulations.

In order to compare to the observations in more detail, a model population was calculated. Each star in the population is assumed to have a 1$M_{Nep}$ planet, a disc with an initial mass and radius randomly selected from the distributions of \cite{wyatt07} and a randomly selected main sequence lifetime from typical main sequence lifetimes for A stars. Ages were selected evenly distributed in log space, as this is consistent with the spread in ages in the observed sample. The smoothed formula for the decrease in accretion rate with time shown in the left-hand panel of Fig.~\ref{fig:posttime} was used to calculate the accretion rates shown in the right-hand panel. Black crosses show our simulations, whilst heavy element accretion rates, calculated from observed Ca abundances \citep{farihi09} are shown as red asterisks. 

Our population agrees qualitatively with these observations, in particular, as also seen in the left-hand panel of Fig.~\ref{fig:posttime}, there is good agreement with the decrease in accretion rate with white dwarf cooling age. There is a surprisingly good agreement with the order of magnitude and range of the observed accretion rates. Our models show that accretion of material onto the star will be a continuous process and for the values of $f_{TD}$ and $f_{acc}$ used in these simulations it is not necessary to invoke discrete events to explain the observations. 

However, our calculated accretion rates scale with the parameters $f_{TD}$ and $f_{acc}$, which will vary between individual planetary systems or disruption events. Our value for the fraction of the disrupted material accreted onto the star ($f_{acc}=0.1$), although reasonable, may well be higher or lower and will vary with factors such as orbital parameters of the body being disrupted, its composition and strength. The largest uncertainty, however, is in the fraction of bodies `scattered in' that end up on star-grazing orbits, $f_{TD}$. This will vary significantly between inner planetary system architectures. White dwarfs with high metal accretion rates that stand out from the population, such as GD362, GD40 and HS 2253+8023, most probably have a planetary system that is particularly efficient at scattering bodies onto star-grazing orbits. Other planetary systems may be less efficient at scattering bodies onto star-grazing orbits, or have this efficiency reduced post-stellar mass loss. In fact, the dynamics of many planetary systems will be altered post-stellar mass loss, potentially inner planets may be scattered such that they collide, are ejected or enter and clear the planetesimal belt. For systems where these processes are relevant, our simple model will no longer apply.

 To determine the importance of further dynamical processes and whether our values for $f_{TD}$ and $f_{acc}$ describe an `average' planetary system, we need to compare the percentage of white dwarfs that are metal polluted found by the observations with our simulations. \cite{Zuckerman10} find that 19\% of DB white dwarfs, with temperatures between 13,500K and 19,500K, are metal polluted with accretion rates $\dot M >10^8$g\,s$^{-1}$. For a comparable sample of DA white dwarfs, \cite{zuckerman03} only found that 5\% had accretion rates $>10^8$g\,s$^{-1}$. These differences may well be attributed to a differences in the birth environments of DA and DB white dwarfs \citep{Zuckerman10}. From our model population in Fig.~\ref{fig:posttime} we calculated a histogram of mass accretion rates, which are shown in Fig.~\ref{fig:mhist} for two age samples, $100$Myr$<t_{cool}<300$Myr and $t_{cool} <10$Gyr.  The former sample corresponds approximately to temperatures betwen 13,500K and 19,500K and 66\% of this sample have $\dot M >10^8$g\,s$^{-1}$, compared to 45\% of the stars with $100$Myr$<t_{cool}<10$Gyr.


These figures suggest that our simulations have overestimated the number of systems for which this simple model is applicable.  The discrepancy could reflect the fraction of planetary systems that are destabilized post-stellar mass loss, since if a planet is scattered into the belt, the amount of material scattered may initially increase, but then decrease at later times as the belt is rapidly cleared. Our simulations thus suggest that either instabilities are relevant for many planetary systems or that most planetary systems, post-stellar mass loss, are significantly less efficient than our solar system at scattering bodies onto star-grazing orbits, i.e. $f_{TD}<0.06$. 

For many main sequence planetary systems our model is too simplistic. It no longer applies if the planet is inclined or eccentric or if the dynamics of the system are dominated by another process, for example secular resonances or binary induced Kozai cycles. Although the evidence for planetesimal belts whose inner edge is truncated by a planet is good, e.g. HR4796 \citep{Wyatt99, Moerchen10}, HD191089 \citep{Churcher10}, Fomalhaut \citep{fomalahautresolvedscatteredlight, chiang_fom}, other explanations have been put forward for inner holes, such as planet formation \citep{KB04, etatel} and interactions with gas \citep{BeslaWu07}.  Hence, the results of these simulations may not apply to all main sequence stars. 

 Our simulations only consider a simplistic model for stellar mass loss. We have assumed spherical symmetry as the natural first assumption, however there have been suggestions that many systems have asymmetric mass loss e.g \cite{soker2001, Parriottalcock}. The stellar mass in the current simulations changed by a factor of 3, when in reality this will vary depending on the initial stellar mass, metallicity and so forth. This potentially changes the total amount of material scattered ($M_{scatt}$) by up to a factor of 2, thus increasing the spread in the calculated accretion rates in Fig.~\ref{fig:posttime}.

We have also ignored other effects of stellar evolution that may cause a decrease in the planetesimal belt mass, for example stellar wind drag, YORP effect or sublimation. \cite{bonsor10} and \cite{Jura2010} show that sublimation has a negligible effect on bodies of purely silicate or mixed composition. \cite{bonsor10} found that stellar wind drag leaves between $10^{-6}$ and $10^{-1}M_{\oplus}$ of material inside of the planetesimal belt at the end of the AGB. These values will be reduced further by the resonance trapping \citep{Dong10}. Nonetheless, this is significantly less than the total mass scattered inwards during our simulations, between $10^{-4}$ and $10^{2}M_{\oplus}$, hence our simulations show that scattering by a planet will dominate over stellar wind drag. 

To summarise, this simple model shows that if every star were to have a planetesimal belt truncated by a planet, and an inner planetary system capable of scattering bodies onto star-grazing orbits, this will produce the observed pollution in white dwarfs. If every system is as efficient as the solar system at scattering bodies onto star-grazing orbits, then a higher fraction of white dwarfs would be metal polluted than is found in observations. Therefore, either many evolved planetary systems are less efficient at scattering bodies onto star-grazing orbits, or further dynamical processes are important.

\section{Conclusions}
\label{sec:conc}


In this work we address the origin of heavy elements in metal polluted white dwarfs and whether accretion of asteroids or comets can explain these observations. We have taken a simple model for a Kuiper-like planetesimal belt whose inner edge is sculpted by a planet. This is typical of many planetary systems on the main sequence. All main sequence stars evolve to become giants, losing a significant proportion of their mass whilst on the asymptotic giant branch to end their lives as a white dwarf (or for higher mass stars than considered here, neutron star or black hole). We have used N-body simulations to investigate the effects of stellar evolution on this simple system, with the focus of explaining observations of metal polluted white dwarfs and white dwarfs with close-in dusty discs. The best models for these systems suggest that they are produced from asteroidal or cometary material that is scattered inwards due to dynamical instabilities post-stellar mass loss. 

We found that for a dynamically cold system ($e_{max}=0$ and $i_{max}=0$), the amount of material scattered in the simulations can be calculated reasonably well using an analytic formulation, shown in Eq.~\ref{eq:manal}. This assumes that pre-mass loss, the chaotic zone, given by Eq.~\ref{eq:chaos}, is cleared, whilst post-stellar mass loss, test particles that are inside of the increased chaotic zone are scattered. For systems with higher initial eccentricities and inclinations, for example $e_{max}=0.1$ and $i_{max}=10^{\circ}$ for the ``cold'' Kuiper belt, the amount of material scattered is higher than given by this analytic formula. The fraction of the belt mass that is scattered increases with planet mass, but is independent of planet semi-major axis. 

Our simulations tracked test particles that are ejected, scattered in, end the simulation in the scattered disc or main belt.  Our definition of `scattered in' included all test particles that are scattered onto orbits with semi-major axis less than $a_{in}=a_{pl}-7r_H$. If there are interior planets, our assumption is that these bodies will interact with them and some fraction will be scattered onto star-grazing orbits, since our simulations show that a single planet is insufficient to scatter bodies onto star-grazing orbits. It is this fraction that is relevant to the white dwarf observations. We found that lower mass planets are more likely to scatter test particles into the inner planetary system, whilst higher mass planets give each particle a larger `kick' with a single close encounter and therefore are more likely to eject them. Hence, despite the increase in the amount of material scattered with planet mass, the mass that is scattered into the inner planetary system is relatively independent of planet mass, given the caveat that the belt mass is not comparable to the planet mass.

In order for our simple model to explain the white dwarf observations enough material must be scattered into the inner planetary system to reproduce the observations. We assume that a fraction $f_{acc} f_{TD}$ of the material `scattered in' is accreted onto the star. Given the wealth of planetary system architectures found in exoplanet systems, for our calculations we take the efficiency of the solar system at scattering Neptune encountering bodies onto sun-colliding orbits ($f_{TD}=0.06$), and an efficiency of the disruption process of $f_{acc}=0.1$. We assume that the initial planetesimal belt properties are the same as those found from observations of debris discs around main sequence A stars, but take into account the collisional evolution of disc material. Accretion rates are calculated using Eq.~\ref{eq:m} and Eq.~\ref{eq:scatt} for a population of evolved planetary systems as a function of their cooling age as a white dwarf (see Fig.~\ref{fig:posttime}). These compare well with the observations, reproducing the correct order of magnitude, approximate range and most importantly the decrease in accretion rate with white dwarf age. Interestingly we find that that stellar mass loss can explain accretion rates even for old ($>1$Gyr) white dwarfs.

 In some ways this agreement is surprising given that the accretion rates scale with $f_{TD}$ and hence will vary significantly between individual planetary systems. In fact our simulations overestimate the number of highly polluted white dwarfs; 82\% of our simulations for white dwarfs with $t_{cool} < 300$Myr have $\dot M > 10^8$\,g\,s$^{-1}$, compared to only 19\% of DB white dwarfs \citep{Zuckerman10} or 5\% of DA white dwarfs \citep{zuckerman03}. There are three factors that could reduce the fraction of white dwarfs with high accretion rates calculated from our simulations. Firstly the efficiency of scattering bodies onto star-grazing orbits may be reduced by the dynamical rearrangement of the planetary orbits. Alternatively, altered dynamics post-stellar mass loss could scatter planets in such a way that they clear the planetesimal belt swiftly of material and hence accretion will not be observed at later times. Finally, not all main sequence stars will have a planetary system resembling this simple model, i.e. containing a planetesimal belt and interior planet on a circular orbit.

This work shows that a simple model with a planetesimal belt and planet is able scatter enough material inwards in order to reproduce the observed metal abundances in polluted white dwarfs, even for old ($>$1Gyr) white dwarfs. In fact, given the observations of debris discs and planets on the main sequence, this model suggests that metals should be observed in a higher proportion of white dwarfs than is found by observations. Either the solar system is particularly efficient at scattering bodies onto star-grazing orbits or dynamical instabilities and the rearrangement of the inner planetary system post-stellar mass loss is crucially important for many evolved planetary systems.

\section*{Acknowledgements}
AB and AJM are grateful to an STFC funded PhD. I (AB) would like to thank Jim Pringle and Jay Farihi for useful discussions and access to the latest data on metal polluted white dwarfs.

\bibliographystyle{mn.bst}

\bibliography{ref}

\end{document}